# Control of photon correlations in type II parametric down-conversion

R. Andrews[1], A. T. Joseph[1], E. R. Pike[2], and Sarben Sarkar[2]

[1] Department of Physics, The University of the West Indies, St.Augustine, Trinidad and Tobago.
Email: randrews@fans.uwi.tt

[2] Department of Physics, King's College London, Strand, London WC2R 2LS, UK.
Email: roy.pike@kcl.ac.uk, sarben.sarkar@kcl.ac.uk

**Abstract**

In this paper we describe theoretically quantum control of temporal correlations of entangled photons produced by collinear type II spontaneous parametric down-conversion. We examine the effect of spectral phase modulation of the signal or idler photons arriving at a 50/50 beam splitter on the temporal shape of the entangled-photon wave packet . The coincidence count rate is calculated analytically for photon pairs in terms of the modulation depth applied to either the signal or idler beam with a spectral phase filter. It is found that the two-photon coincidence rate can be controlled by varying the modulation depth of the spectral filter.

1. **Introduction**

The optical process of spontaneous parametric down-conversion (SPDC) involves the virtual absorption and spontaneous splitting of an incident (pump) photon in a transparent nonlinear crystal producing two lower-frequency (signal and idler) photons [1-3]. The pairs of photons can be entangled in a multi-parameter space of frequency, momentum and polarization. In type-II SPDC the photons are entangled in frequency and wave number and signal and idler photons have orthogonal polarizations. In noncollinear type-II SPDC the signal and idler photons are entangled as well in polarization but in the collinear case there is no polarization entanglement [3]. Entangled photons have been used to demonstrate quantum nonlocality [4,5], quantum teleportation [6-8] and, more recently, quantum information processing [9-11] and quantum cryptography [12,13]. Recently there has been much interest in controlling the  temporal properties of entangled photons and this has been achieved either by spectral filtering with narrowband filters [14,15], or by placing the nonlinear crystal in cavities [16,17]. Temporal shaping has also been obtained more recently using spectral phase modulation [18] of signal or idler photons produced in type I collinear parametric down-conversion.

In contrast to previous work, we present a detailed theoretical model for describing the temporal shaping of  spectrally phase-modulated photons in collinear type-II parametric down-conversion. The two-photon wave packet produced in type-II parametric down-conversion has already been measured in a simple beam-splitting experiment [19]. By placing birefringent material between the BBO ($\beta$- $BaB_2O_2$ ) crystal and the beam splitter an optical delay is introduced between the signal and idler photons and this leads to a V-shaped coincidence count rate. Here we obtain the temporal shape of the two-photon wave packet when either the signal or idler photons are phase modulated before the detection process.

## 2. Theoretical Model

The amplitude for detecting photon pairs at conjugate space-time points $(\vec{r}_1, t_1)$ and $(\vec{r}_2, t_2)$ is given by

$$A^{(2)} = \langle \vec{E}_H^+(\vec{r}_1, t_1) \vec{E}_H^+(\vec{r}_2, t_2) \rangle \tag{1}$$

where $t_1$ and $t_2$ are detection times of signal and idler photons and $\vec{E}_H(\vec{r}_i, t_i)$ ($i=1,2$) are the Heisenberg electric field operators [20]. In the steady state the right-hand-side of (1) can be expressed as

$$A^{(2)} \propto \int d^3 r_3 \int d^3 k_1 \int d^3 k_2 U^*_{\vec{k}_1 \lambda_1}(\vec{r}_3) U^*_{\vec{k}_2 \lambda_2}(\vec{r}_3) U_{\vec{k}_0 \lambda_0}(\vec{r}_3) f_p(\vec{r}_3) \left(\frac{\hbar \omega_{k_0}}{2\varepsilon_0}\right)^{\frac{1}{2}} \left(\frac{\hbar \omega_{k_1}}{2\varepsilon_0}\right)^{\frac{1}{2}} \left(\frac{\hbar \omega_{k_2}}{2\varepsilon_0}\right)^{\frac{1}{2}}$$

$$\times \langle \alpha_{k_0}, 0 | \vec{E}_I^{(+)}(\vec{r}_2, t_2) \vec{E}_I^{(+)}(\vec{r}_1, t_1) | \alpha_{k_0}, k_1, k_2 \rangle \delta(\omega_{k_1} + \omega_{k_2} - \omega_{k_0}) \tag{2}$$

where $\vec{E}_I(\vec{r}_i, t_i)$ ($i=1,2$) are the interaction-picture electric field operators at some arbitrary but specified detection points $\vec{r}_1$ and $\vec{r}_2$ and $f_p(\vec{r}_3)$ is a function which describes the shape of the pump in the transverse direction; $U_{\vec{k}_i \lambda_i}(\vec{r}_3)$ ($i=0,1,2$) are plane-wave modes describing the electromagnetic field in free space with $\vec{k}_1, \vec{k}_2$ as the wave vectors of the signal and idler photons, $\vec{k}_0$ is the wave vector of the incident pump photon and $\lambda_i$ are polarization indices. (As in our previous studies we have not introduced the effects of a change in the linear refractive index between the crystal and its surroundings. The incorporation of this difference does not qualitatively change our conclusions.) The initial state of the electromagnetic field $|0, \alpha_{\vec{k}_0}\rangle$, consists of a coherent state with wave-vector $\vec{k}_0$ and frequency $\omega_{k_0}$ (the monochromatic pump beam) with other modes in the vacuum state $|0\rangle$. Now consider the following experiment in Figure 1 in which a crystal is pumped by a narrow-band cw laser and *o*- and *e*-polarized photons propagate to a beam splitter, BS, and are then detected in coincidence at detectors D1 and D2 located at $\vec{r}_1$ and $\vec{r}_2$ respectively. Belinsky and Klyshko [23] have shown that photon correlation after the beam splitter is determined by the Fourier spectrum of photons produced in the parametric process. In this type of experiment the photon pairs become entangled in polarization [24]. A variable optical delay $\tau$ is given to one of the photons of a pair (say the extraordinary photon or *e*-photon) and the coincidence count rate is measured as a function of the relative delay $\tau$. The electric fields at the detectors D1 and D2 are given by (see eqn. (14) in [21])

$$\vec{E}_I^+(\vec{r}_1, t_1) = \frac{1}{\sqrt{2}} \left[\vec{E}_{I,o}^+(\vec{r}_1, t_1) + \vec{E}_{I,e}^+(\vec{r}_1, t_1 + \tau)\right]$$

$$\vec{E}_I^+(\vec{r}_2, t_2) = \frac{1}{\sqrt{2}} \left[\vec{E}_{I,o}^+(\vec{r}_2, t_2) - \vec{E}_{I,e}^+(\vec{r}_2, t_2 + \tau)\right] \tag{3}$$

where $E^+_{I,o}$ and $E^+_{I,e}$ are the electric fields for ordinary and extraordinary photons respectively. The interaction-picture quantized electric field $\vec{E}^{(+)}_I(\vec{r},t)$ is given as

$$\vec{E}^+_I(\vec{r},t) = i\sum_\lambda \int d^3k \left(\frac{\hbar\omega_k}{2\varepsilon_0}\right)^{1/2} \vec{\varepsilon}_{\vec{k}\lambda} e^{i(\vec{k}\cdot\vec{r}-\omega t)} a_{\vec{k}\lambda} \quad (4)$$

where $a_{\vec{k}\lambda}$ is the annihilation operator for photons with wave vector $\vec{k}$ and $\lambda$ is the polarization index; $\varepsilon_\lambda(\vec{k})$ ($\lambda = 1,2$) denotes the mode polarization vector. We assume that the divergence of the pump is negligible over the length of the crystal and that the transverse shape of the pump is Gaussian, i.e., we take $f_p(\vec{r}_3)$ as

$$f_p(\vec{r}_3) \propto \exp\left(-\frac{r_{3x}^2 + r_{3y}^2}{\varepsilon_\perp^2}\right) \quad (5)$$

The plane-wave mode functions in (2) are given as $U_{\vec{k}_i\lambda_i}(\vec{r}_3) = \exp(-i\vec{k}_i \cdot \vec{r}_3)$. After performing the $\vec{r}_3$-integration in (2) and using the transformations in (3) we obtain

$$A^{(2)} \propto \iint d\omega_{k_1} d\omega_{k_2} \mathrm{sinc}\left(\Delta k \frac{d}{2}\right)\left[e^{i(\vec{k}_1\cdot\vec{r}_2 - \omega_{k_1} t_2)} e^{i(\vec{k}_2\cdot\vec{r}_1 - \omega_{k_2}(t_1+\tau))} - e^{i(\vec{k}_1\cdot\vec{r}_1 - \omega_{k_1} t_1)} e^{i(\vec{k}_2\cdot\vec{r}_2 - \omega_{k_2}(t_2+\tau))}\right]\delta(\omega_{k_1} + \omega_{k_2} - \omega_{k_0}) \quad (6)$$

where $\vec{k}_1, \vec{k}_2$ are the wave vectors of the ordinary and extraordinary photons respectively, $d$ is the length of the nonlinear crystal and $\Delta k = k_0 - k_1 - k_2$. The two-photon state has a finite bandwidth so we let $\omega_{k_1} = \omega_{k_1^*} + \nu$ and $\omega_{k_2} = \omega_{k_2^*} - \nu$ where $|\nu| \ll \omega_{k_{1,2}^*}$ and $\omega_{k_{1,2}^*}$ are phase-matched frequencies of the signal and idler photons. Expanding $k_1$ and $k_2$ to first order in $\nu$ we obtain

$$k_1 = k_1^* + \nu/u_o$$
$$k_2 = k_2^* - \nu/u_e \quad (7)$$

where $u_o$ ($u_e$) is the group velocity for the ordinary (extraordinary) photons. If we consider the degenerate case in which $\omega_{k_{1,2}^*} = \frac{\omega_{k_0}}{2}$ and use the delta function in (6) together with the dispersion relations for the wave numbers in (7) we obtain, after integrating over all $t_1$ and $t_2$, the modulus square of $A^{(2)}$ proportional to the following frequency integral:

$$\left|A^{(2)}\right|^2 \propto \int d\nu \, \mathrm{sinc}^2[\tau_1\nu](1 - \cos(2\nu[\tau+\tau_2])) \quad (8)$$

where

$$\tau_1 = \left(\frac{1}{u_0} - \frac{1}{u_e}\right)\frac{d}{2} \text{ and } \tau_2 = \left(\frac{1}{u_0} - \frac{1}{u_e}\right)z \quad (9)$$

and we have assumed that each detector is of equal distance $z$ from the centre of the nonlinear crystal. Figure 2 gives a plot of $\left|A^{(2)}\right|^2$ against an effective time delay $\tau' = 0.14 \times 10^{14}(\tau + \tau_2)$ for $\left(\dfrac{1}{u_0} - \dfrac{1}{u_e}\right) = 2.5$ ps/cm and $d = 0.56$ mm. From Figure 2 the estimated size of the half-base of the triangle, $\Delta\tau' \approx 1$ which gives $\Delta\tau \approx 72$ fs in good agreement with the results in [19].

## 3. Phase-modulated entangled photons

We now analyse the same experiment in Figure 1 but in addition we phase-modulate the o-polarized photons using a spectral phase filter. The quantized electric field describing such phase-modulated photons is given as

$$\vec{E}_I^+(\vec{r},t) = i\sum_\lambda \int d^3k \left(\dfrac{\hbar\omega_k}{2\varepsilon_0}\right)^{1/2} \vec{\varepsilon}_{\vec{k}\lambda} e^{i(\vec{k}\cdot\vec{r}-\omega_k t)} e^{i\theta(\omega_k)} a_{\vec{k}\lambda} \qquad (10)$$

where $\theta(\omega_k)$ is the spectral phase distribution introduced by the spectral phase filter. If we consider a symmetric spectral phase distribution of the form $\theta(\omega_k) = \alpha\cos(\beta\omega_k)$ [22] where $\alpha$ and $\beta$ are the modulation depth and frequency respectively, we obtain $|A^{(2)}|^2$ proportional to the following frequency integral, i.e.,

$$|A^{(2)}|^2 \propto \int d\nu \, \text{sinc}^2[\tau_1\nu]\left(2 - \left\{e^{-i(2\nu[\tau+\tau_2])} e^{i2\alpha\sin\left(\dfrac{\beta\omega_{k_0}}{2}\right)\sin(\beta\nu)} + c.c\right\}\right) \qquad (11)$$

The above integral can be rewritten in the form

$$|A^{(2)}|^2 \propto \int d\nu \, \text{sinc}^2[\tau_1\nu]\Bigg(1 - J_0(\gamma)\cos(2\nu[\tau+\tau_2]) - 2\sum_{n=1}^\infty J_{2n}(\gamma)\cos(2n\beta\nu)\cos(2\nu[\tau+\tau_2])$$

$$- 2\sum_{n=1}^\infty J_{2n-1}(\gamma)\sin([2n-1]\beta\nu)\sin(2\nu[\tau+\tau_2])\Bigg) \qquad (12)$$

where $J_n$ is a Bessel function of order $n$ and $\gamma = 2\alpha\sin\left(\dfrac{\beta\omega_{k_0}}{2}\right)$. The experiment we describe is such that the degenerate photons are of wavelength 700 nm and $\beta = 50$ fs. We also take $\left(\dfrac{1}{u_0} - \dfrac{1}{u_e}\right) = 2.5$ ps/cm and the crystal length $d = 0.56$ mm. These values give $\tau_1 = 7\times10^{-14}$ s and $\tau_2 = 1.3\times10^{-10}$ s. We consider coincidence counts for a fixed effective optical delay $\tau' = 0$, where $\tau' = 2\times10^{14}(\tau+\tau_2)$, as we vary the modulation depth $\alpha$. Figure 3 is a plot of $|A^{(2)}|^2$ against $\gamma$. Figure 3 shows that the variation of the coincidence counts against $\gamma$ for photons that have been given an optical delay $\tau = -\tau_2 = -13$ ns. To observe the effect of spectral phase modulation on the temporal shape of the two-photon wavepacket we plot of $|A^{(2)}|^2$ against $\tau'$ (Figure 4) for $\gamma = 4$ and $\gamma = 7$ with $\beta = 50$ fs. Figure 4 clearly shows that the position of maximum count rate can be controlled by varying the modulation depth. It is therefore possible to maximize the number of pairs arriving with zero optical delay at an atom. This can be used to enhance the two-photon absorption rate which requires two photons, whose sum energy equals that of the atomic transition, to arrive at the atom together.

## 4. Conclusions

We have analyzed theoretically the temporal shape of two-photon correlations in type-II parametric down-conversion. We have shown that it is possible to optimize the number of photon pairs arriving simultaneously at an absorbing atom. This can be achieved by placing a symmetric spectral phase filter with a variable modulation depth in the path of one photon of a pair. Since it is important in two-photon absorption studies (e.g. microscopy and spectroscopy) for photons to arrive in pairs simultaneously at target molecules we believe our work would be valuable in optimizing two-photon interactions particularly in the low flux limit. In addition it would be interesting to consider other types of phase-modulated pulses as well as quantum interference effects with temporally distinguishable pulses [25].

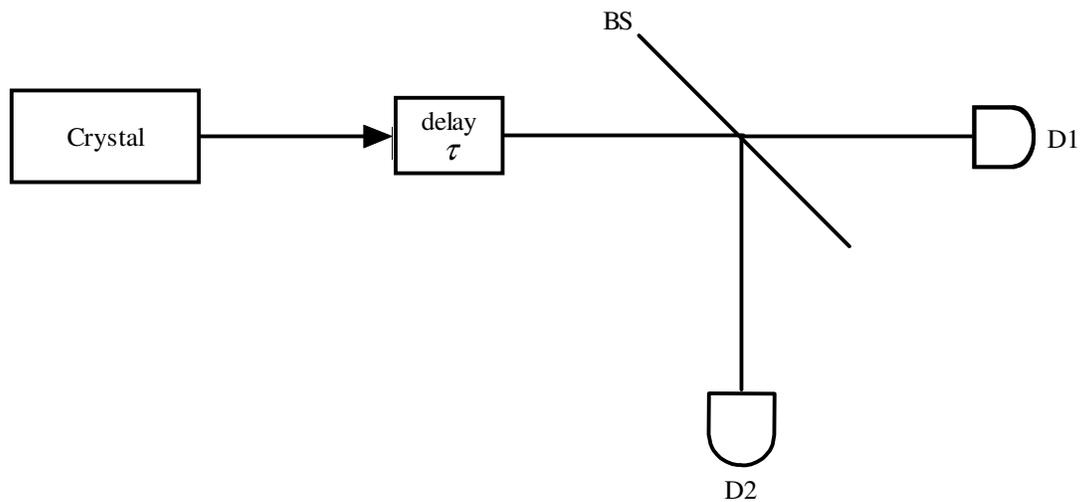

Figure 1: Schematic set-up of an experiment in which a type-II crystal is pumped by a narrow-band cw laser producing *o*- and *e*-polarized photons which propagate to a beam splitter (BS) and are detected in coincidence at detectors D1 and D2; $\tau$ is the optical delay given to the *e*-polarized photon.

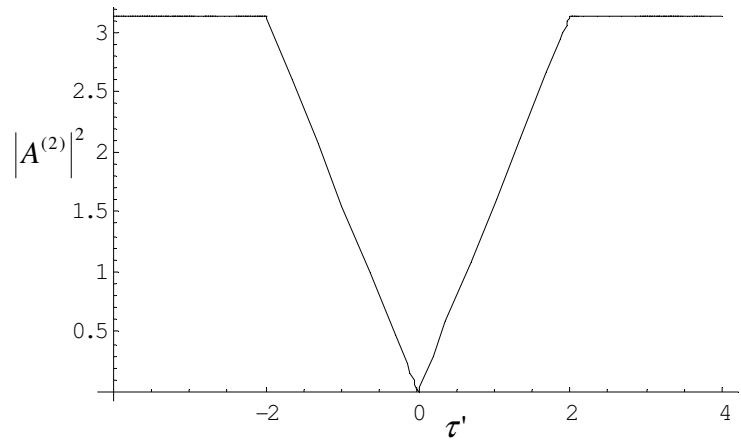

Figure 2: Plot of the coincidence count rate $\left|A^{(2)}\right|^2$ (arbitrary units) against the effective optical delay $\tau' = 0.14 \times 10^{14} (\tau + \tau_2)$

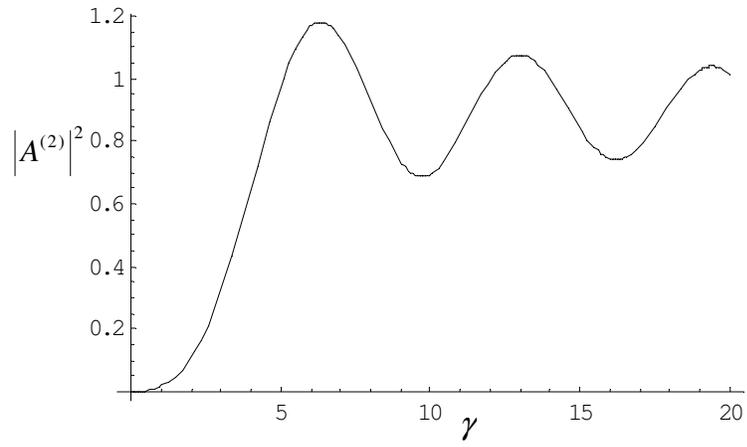

Figure 3: Plot of the coincidence count rate $\left|A^{(2)}\right|^2$ against $\gamma$ for degenerate photons of wavelength 700 nm with $\beta = 50$ fs and a crystal of length 0.56 mm.

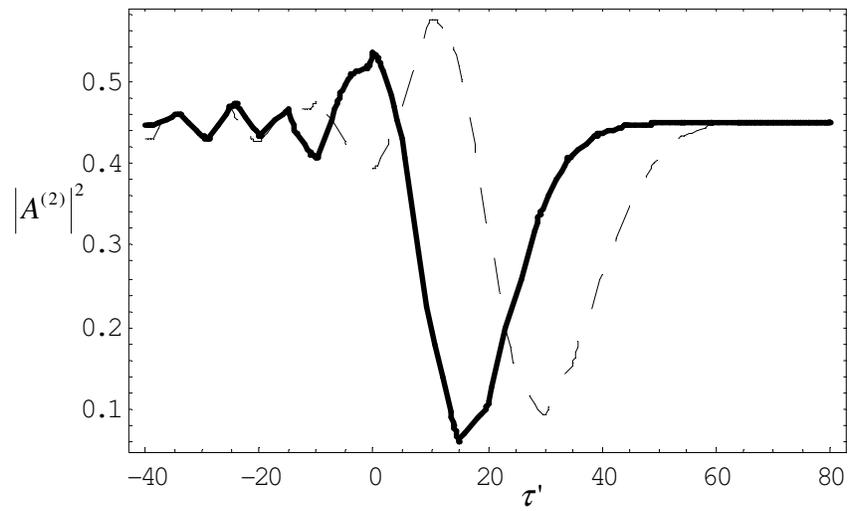

Figure 4: Plot of the coincidence count rate against $\tau'$ for degenerate photons of wavelength 700 nm with $\beta = 50$ fs for $\gamma = 4$ (solid curve) and $\gamma = 7$ (dashed curve).